\documentclass{cpbtex}

\usepackage{graphicx}
\usepackage{booktabs}
\usepackage{caption}
\usepackage{subcaption}
\usepackage{amsmath}
\usepackage{extarrows}
\usepackage{multirow}
\usepackage{threeparttable} 

\begin{document}
\begin{CJK*}{GBK}{song}

\title{Exploring fundamental laws of classical mechanics via predicting the orbits of planets based on neural networks\thanks{Project supported by the National Natural Science Foundation of China (Grant No.11975050).}}

\author{Jian Zhang, \ Yiming Liu \ and \ Z.C. Tu\thanks{Corresponding author. E-mail:~tuzc@bnu.edu.cn}\\
{Department of Physics, Beijing Normal University, Beijing 100875, China}} 

\date{\today}
\maketitle

\begin{abstract}
Neural networks have provided powerful approaches to solve various scientific problems. Many of them are even difficult for human experts who are good at accessing the physical laws from experimental data. We investigate whether neural networks can assist us in exploring the fundamental laws of classical mechanics from data of planetary motion. Firstly, we predict the orbits of  planets in the geocentric system using the gate recurrent unit, one of the common neural networks. We find that the precision of the prediction is obviously improved when the information of the Sun is included in the training set. This result implies that the Sun is particularly important in the geocentric system without any prior knowledge, which inspires us to gain Copernicus' heliocentric theory. Secondly, we turn to the heliocentric system and  make successfully mutual predictions between the position and velocity of planets. We hold that the successful prediction is due to the existence of enough conserved quantities (such as conservations of mechanical energy and angular momentum) in the system. Our research provides a new way to explore the existence of conserved quantities in mechanics system  based on neural networks.

\end{abstract}



\section{\label{sec:level1}Introduction}
Exploration of physical laws through the analysis of experimental data extremely promoted to the development of physics in history. People had generally believed in the geocentric theory for a long time until the 16th century. Copernicus proposed the heliocentric theory which corrected previous misconceptions of human view on the world. The heliocentric theory plays an important role in the history of physics and astronomy \cite{nature1943}. Based on the heliocentric theory, Kepler discovered three laws of planetary motion through analyzing the experimental data observed by Tycho Brahe. Newton proposed three laws of motion and the law of universal gravitation, and further provided a good explanation for Kepler's laws of planetary motion. From then on, the framework of classical mechanics was established. The journey from the geocentric theory to the heliocentric theory, and then to Newton's mechanics represents a huge leap for exploring the laws of nature by human intelligence. 

Nowadays, with the flourishing development of artificial intelligence (AI) \cite{nature2015,DL}, machine learning has become a new ``smart" tool for analyzing experimental data in fundamental science, such as astrophysics \cite{Caldeira2019,Yu}, biological physics \cite{Pfeil,Pet}, condensed-matter physics \cite{zhaihui,ma2021,kau,sturm,yichen,ZhaoR}, engineering mechanics\cite{ma2020}, high-energy physics \cite{baldi2016}, statistical physics \cite{been,Giri,Rotondo,huang,casert,Liujg} and so on. It is the latest trend to discover fundamental laws of physics based on machine learning without prior experience about physics. Zhao \cite{zhao} inferred the dynamics of  two models including a representative model of low-dimensional nonlinear dynamical systems and a spatiotemporal model of reaction-diffusion systems. He found that, instead of establishing equations of motion, the learning machine could be uesd to infer the dynamic properties of  ``black-box'' systems. Iten et al. \cite{Iten2020} investigated a toy system consisting of the Sun, Mars and Earth by building a neural network named SciNet.  Although the training data only contain angles of Mars and the Sun as seen from Earth, they found that the SciNet could switch these angles to a heliocentric representation. This finding implies that  the SciNet can gain conceptual insight that the solar system is heliocentric. Qin \cite{qin2020} predicted the orbits of planets in the solar system by designing a method for learning and serving. It is found that  the algorithm could learn Kepler's laws to some extent. In view of these works, we believe that neural networks will bring infinite possibilities for research. In this paper, we employ gated recurrent unit (GRU) \cite{gru1,gru2,gru3}, one of the common neural networks, to explore the fundamental laws of classical mechanics via predicting the orbits of planets. Although the training data only contain position information of planets and the Sun in geocentric system, we find that the Sun is crucial to the generalization ability of neural networks, which inspires an insight that the solar system is heliocentric. We further obtain the evidence for the existence of conserved quantities by making mutual predictions between position and velocity of planets in heliocentric system. The rest content of our paper is organized as follows. In Sec.~\pageref{sec:level2}, we introduce the data of planetary position and velocity and structure of the neural network we adpot. In Sec.~\pageref{sec:level3}, we present our results on the prediction of planetary orbits in geocentric system. In Sec.~\pageref{sec:level4}, we explore the existence of conserved quantities  in mechanical systems. The last section is a brief summary and discussion.

\section{\label{sec:level2}Data set and structure of neural network}

The initial  data of planetary position and velocity in the heliocentric system come from Jet Propulsion Laboratory  Planetary and Lunar Ephemeris DE421 \cite{de421}. We obtain time-dependent data sets of planetary position and velocity in the heliocentric system by solving a set of differential equations  governed by Newton's second law and the law of universal gravitation with time interval 0.01 years (Earth years). Here, we merely consider gravitation between the Sun and planets. The data of planets and the Sun in geocentric system are obtained by subtracting the data of Earth from the data in heliocentric system. We regard these data as true values of ``experimental" data.

\begin{figure}[ht]
	\centering	
	\subcaptionbox{\label{subfig:1a}}
	{
		\includegraphics[width=7cm]{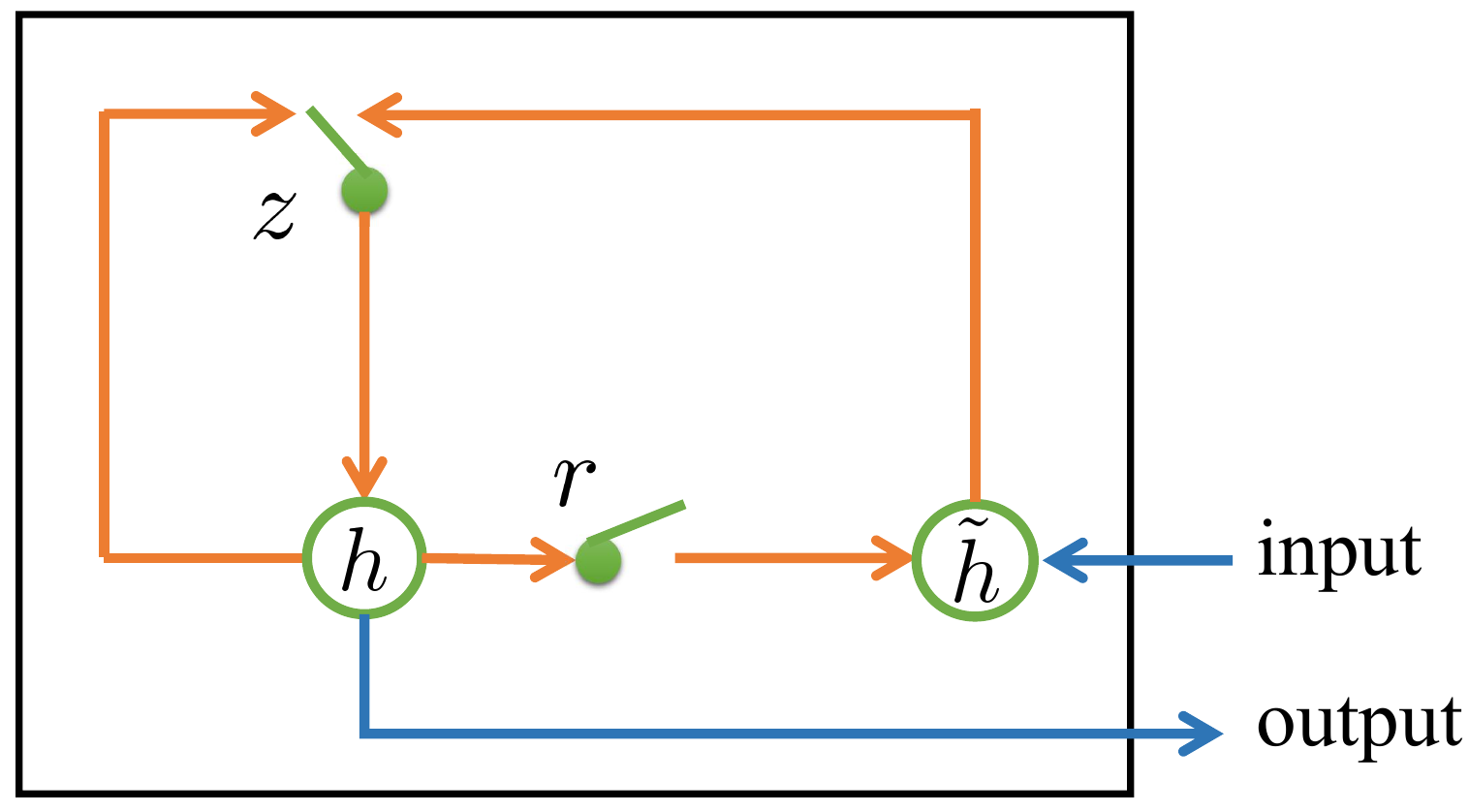}
	}	
	\subcaptionbox{\label{subfig:1b}}
	{
		\includegraphics[width=7cm]{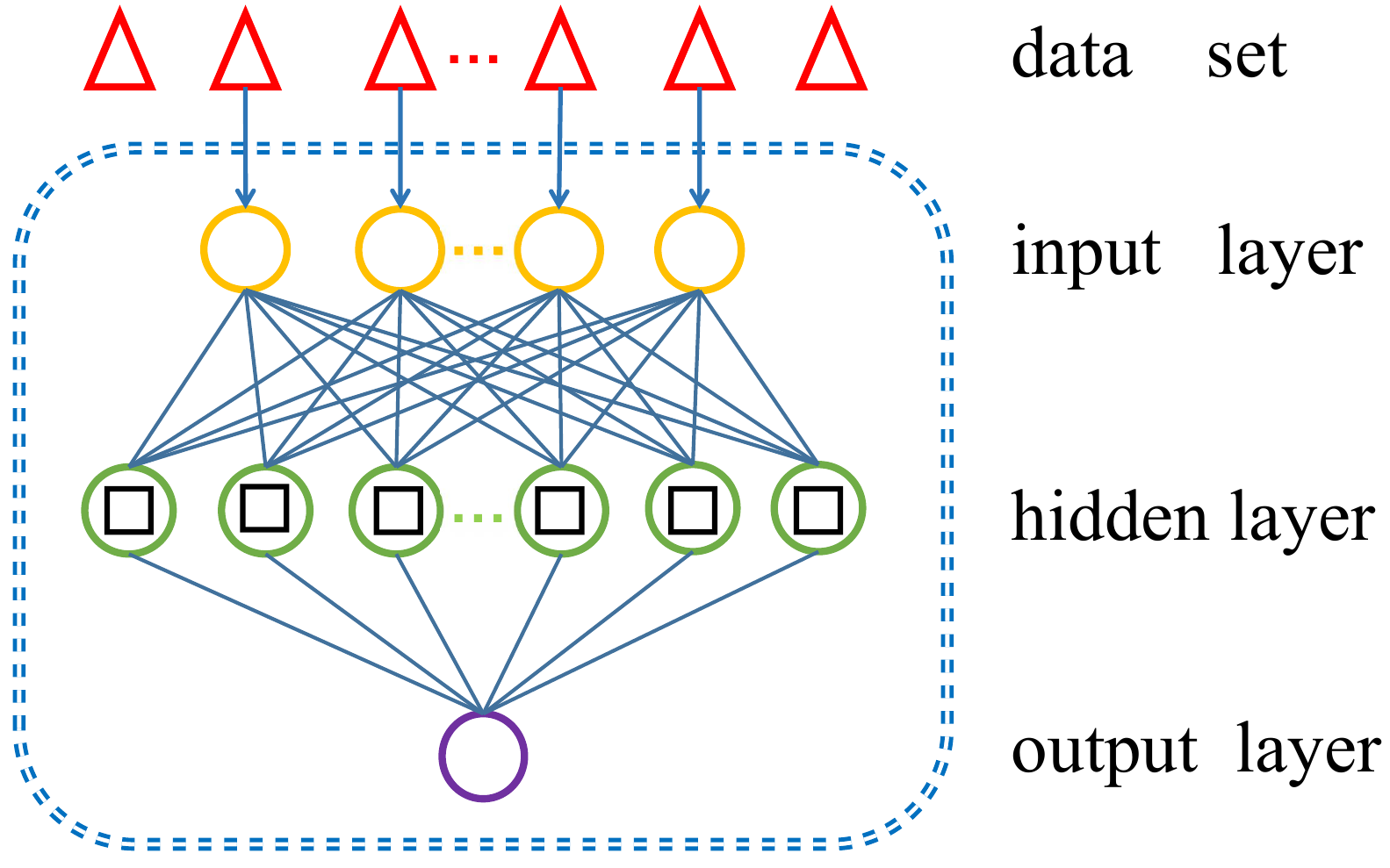}
	}
	\caption{(Color online) Graphical illustration of the neural network. (a) gated recurrent unit, which includes the reset gate $r$, the update gate $z$, the activation $h$ and the candidate activation $\tilde{h}$ \cite{gru1,gru2}. (b) the structure of neural network we use, which consists of an input layer, a hidden layer and an output layer. Note that (a) is the concrete structure of each hidden neuron in (b).}
	\label{structure}
\end{figure}

Recurrent neural network \cite{rnn1} has been widely used in processing time-series beacuse of its powerful fitting ability \cite{rnn2}. GRU is an improved model of recurrent neural network as shown in Fig.~\ref{subfig:1a}. The main characteristic for GRU \cite{gru1,gru2} is controlling the update of information by introducing gating units, which include a reset gate $r$, an update gate $z$, an activation $h$ and a candidate activation $\tilde{h}$. The structure of the neural network in this work is shown in Fig.~\ref{subfig:1b}. The neural network contains one hidden layer, which is composed of GRU and fully connected with an input layer and an output layer. We use supervised learning algorithm and  carry out our work with Keras \cite{keras} running on top of Tensorflow \cite{tf}.

\section{\label{sec:level3}Prediction of planetary orbits in geocentric system}

In this section, we use the neural network to predict the orbits of planets which consist of Venus, Mars and Jupiter. The detailed hyperparameters of the neural network are shown in Table~\ref{table1}. The input layer and output layer contain 10 neurons and 1 neuron, respectively. The training set contains 14-year data, totally 1400 samples, and the test set contains 4-year data, totally 400 samples. 

\begin{table*}[ht]
	\centering
	\footnotesize
	\caption{Hyperparameters of the neural network in Section 3}\label{table1} 
	\begin{tabular}{ccccccc}
	    \toprule
		Loss function&Hidden units&Input units&Epoch&Learning rate&Batch size&Optimizer \\
		\midrule
		Mean squared error&600&10&50&0.2&8 & AdaGrad\cite{adagrad} \\
		\bottomrule
	\end{tabular}
\end{table*}

The data processing is as below. We train the neural network to learn a mapping $f$:
\begin{equation}
	\vec{U}_{n+10} = f(\vec{U}_{n},\vec{U}_{n+1},\vec{U}_{n+2},\dots,\vec{U}_{n+9}),~n=1,2,3\dots	\label{f}
\end{equation}
through the training set, where $\vec{U}_{n}$ is an element of time-series. Then we use the trained neural network to predict planetary orbits. During the training, all input $\vec{U}_{n}$ ($n$=1 to 1400) are true values in training set. During the prediction, we merely input ten true values of $\vec{U}_{n}$ ($n$= 1401 to 1410) in test set. And then we output  the  predicted  values of $\vec{U}_{n}$~($n$=1411 to 1800). 

The neural network operates on two types of data in geocentric system. One merely includes the position information of a planet, i.e., $\vec{U}_{n}$=$(x_{n}^a,y_{n}^a,z_{n}^a)$, where $x_{n}^a$, $y_{n}^a$ and $z_{n}^a$ are the coordinates of the planet in geocentric system. Another includes simultaneously position information of a planet and the Sun in geocentric system, i.e.,  $\vec{U}_{n}$=$(x_{n}^a,y_{n}^a,z_{n}^a,x_{n}^s,y_{n}^s,z_{n}^s)$, where $x_{n}^s$, $y_{n}^s$ and $z_{n}^s$ are the coordinates of the Sun. We normalize the data with  constant $X_{max}-X_{min}$, where $X_{max}$ and $X_{min}$ are the maximum and the minimum of $x_{n}^a $ for n=1 to 1400, respectively. For example, $y_{n}^a $ is transformed into $ ({y_{n}^a-X_{min}})/({X_{max}-X_{min}}) $.

\begin{figure}
	\centering
	\subcaptionbox{\label{subfig:2a}}
	{
		\includegraphics[width=7cm]{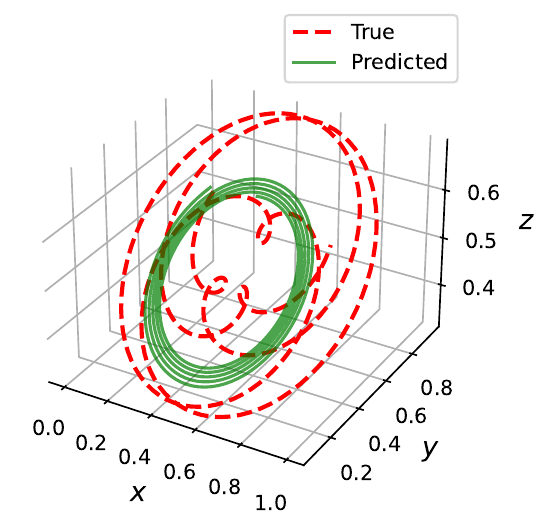}\quad 
		\includegraphics[width=7cm]{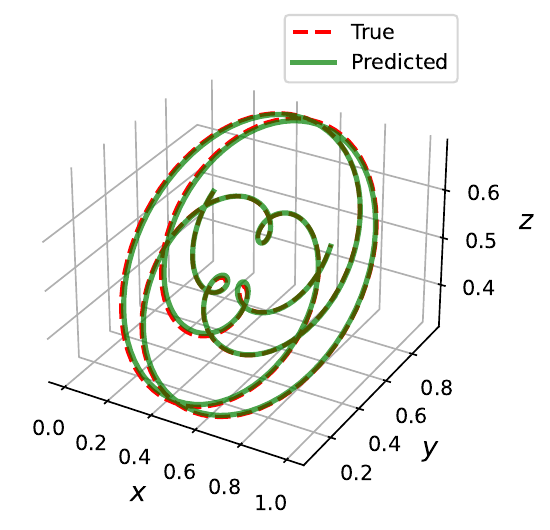}
	}
	\subcaptionbox{\label{subfig:2b}}
	{
		\includegraphics[width=7cm]{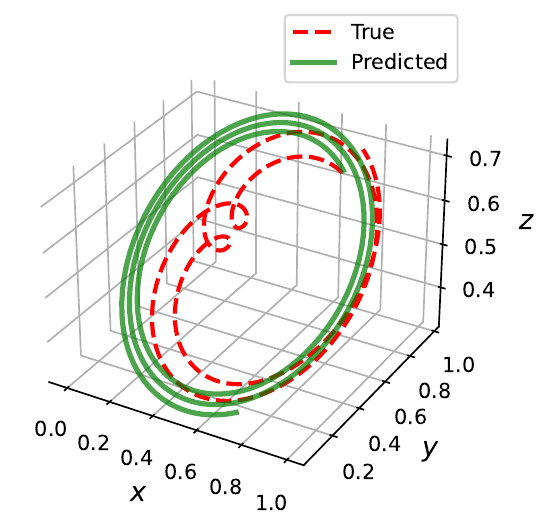}\quad 
		\includegraphics[width=7cm]{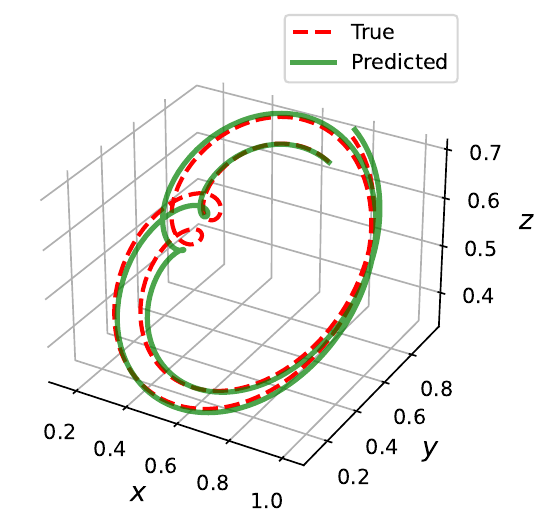}
	}
	\subcaptionbox{\label{subfig:2c}}
	{
		\includegraphics[width=7cm]{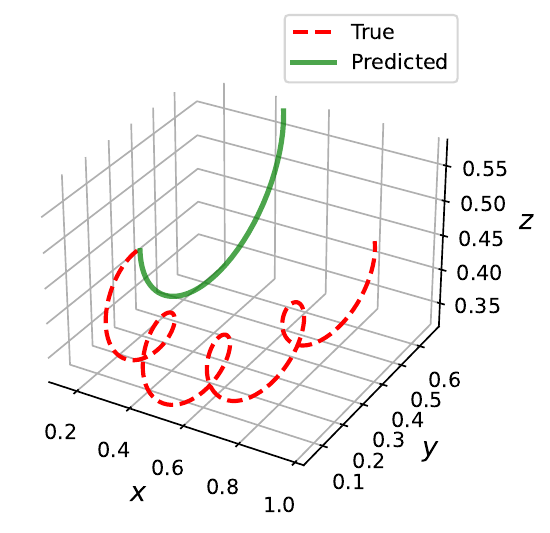}\quad 
		\includegraphics[width=7cm]{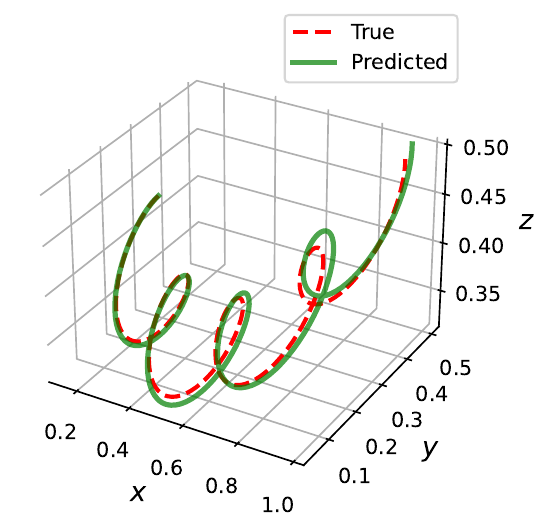}
	}
	\caption{(Color online) Images of planet orbits. The dashed line represents the true values, and the solid line represents the predicted values. The results without considering information of the Sun is on the left, and the results with information of the Sun is on the right. (a) Orbit of Venus. (b) Orbit of Mars. (c) Orbit of Jupiter.}
	\label{fig2}
\end{figure}

The predicted orbits of Venus, Mars and Jupiter are shown in Fig.~\ref{fig2}. Each figure contains two curves. The dashed  line represents true orbit, and the solid line represents predicted orbit. We show the result without considering information of the Sun in the left column. And we show the result with information of the Sun in the right column. By glancing at these images, we find that the predicted values are closer to the true values when considering the Sun. As shown in Table \ref{tableerror1}, we  calculate the test errors (mean squared error)  by comparing the predicted values and the true values. The  precision for predicted cruves has been significantly improved  with the information of the Sun.

\begin{table}[h]
	\centering
	\caption{Test errors of  prediction for planetary orbits in Fig.~\ref{fig2} }\label{tableerror1}  
	\begin{threeparttable}  
		\begin{tabular}{*{7}{c}}
			\bottomrule
			\multirow{2}*{Planet}&\multicolumn{2}{c}{Venus}&\multicolumn{2}{c}{Mars}&\multicolumn{2}{c}{Jupiter} \\
			\cmidrule(lr){2-3}\cmidrule(lr){4-5}\cmidrule(lr){6-7}
			&left\tnote{1}&right\tnote{2}&left\tnote{1}&right\tnote{2}&left\tnote{1}&right\tnote{2}\\
			\midrule
			Test error &2.8e-04&8.5e-08&3.3e-05&2.1e-07& 4.6e-05&1.1e-08\\
			\bottomrule
		\end{tabular}
		\begin{tablenotes}    
			\footnotesize              
			\item[1] without considering information of the Sun.    
			\item[2] with information of the Sun.        
		\end{tablenotes}            
	\end{threeparttable}       
\end{table}

\begin{figure}[ht]
	\centering
	\subcaptionbox{\label{subfig:3a}}
	{
		\includegraphics[width=7cm]{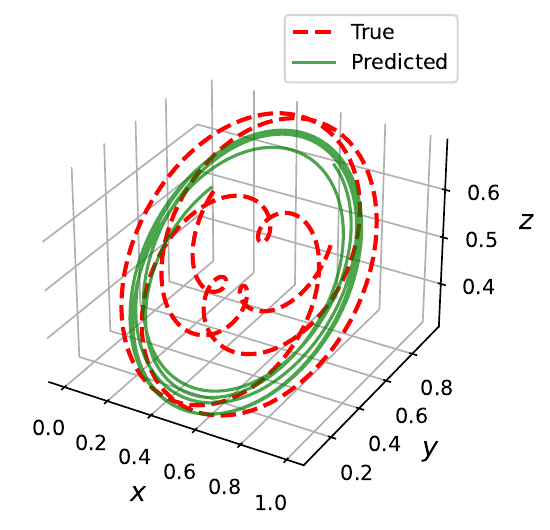}
	}
	\subcaptionbox{\label{subfig:3b}}
	{
		\includegraphics[width=7cm]{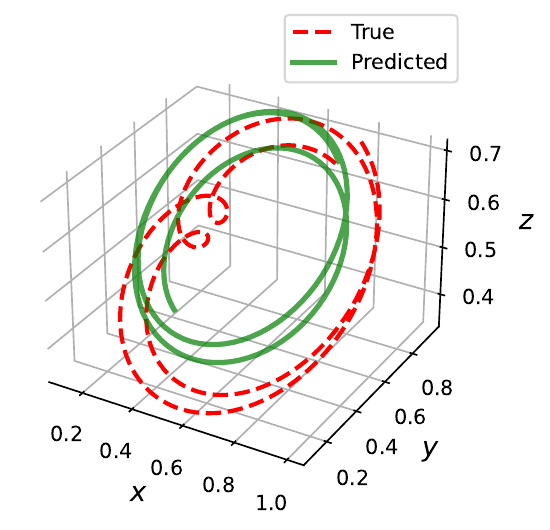} 
	}
	\caption{(Color online) Images of orbits with information of Jupiter rather than the Sun. (a) Orbit of Venus. (b) Orbit of Mars.}
	\label{fig3}
\end{figure}

We also  predict orbits of planets with information of another planet rather than the Sun. We predict orbits of Venus and Mars with information of Jupiter. As shown in Fig.~\ref{fig3}, the precision of prediction decreases obviously. Without any prior knowledge, these facts imply that the Sun is particularly important in the model system including Venus, Mars, Jupiter, Earth and the Sun. In a way, this finding based on neural networks helps us to gain Copernicus' heliocentric theory. This is one of the main results in our paper.

\section{\label{sec:level4}Evidence for the existence of conserved quantities}

According to the universal approximation theorem \cite{cybenko1989,hornik1989}, the reason for successful prediction is that neural networks can learn a mapping from big data. The predicted results of neural networks may indicate whether there are conserved quantities in a mechanical system. We propose that if there are enough conserved quantities in a system, neural networks can successfully learn the mapping between positon and velocity. In order to explore the existence of conserved quantities, we turn to a model system including Venus, Mars and Jupiter in heliocentric system since  we have found that the Sun plays a very important role in predicting orbits in the previous section. 

We adopt the same structure of neural network shown in Fig.~\ref{structure}. Unlike in the previous section, here the input layer and output layer only contain one neuron, respectively. The training set contains 14-year data of position and velocity in heliocentric system, totally 1400 samples, and the test set contains 12-year data, totally 1200 samples. The hyperparameters of the neural network are shown in Table~\ref{table2}.

\begin{table*}[ht]
	\centering
	\caption{Hyperparameters of the neural network in Section 4}\label{table2} 
	\begin{tabular}{ccccccc}
		\toprule
		Loss function&Hidden units&Input units&Epoch&Learn rate&Batch size&Optimizer \\
		\midrule
		Mean squared error&90&1&50&0.2&8 & AdaGrad\cite{adagrad} \\
		\bottomrule
	\end{tabular}
\end{table*}

\begin{figure}
	\centering
	\subcaptionbox{\label{subfig:4a}}
	{
		\includegraphics[width=7cm]{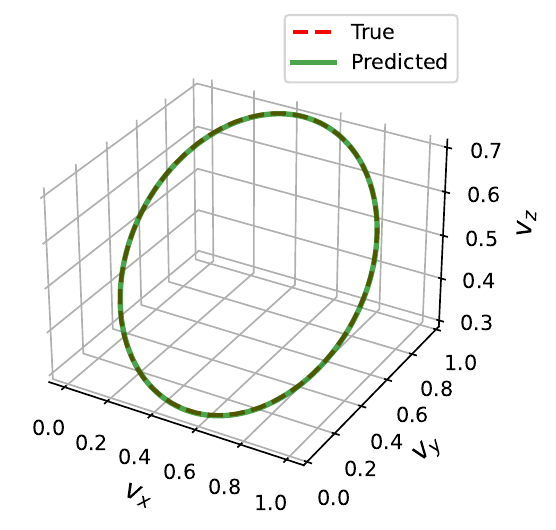}\quad 
		\includegraphics[width=7cm]{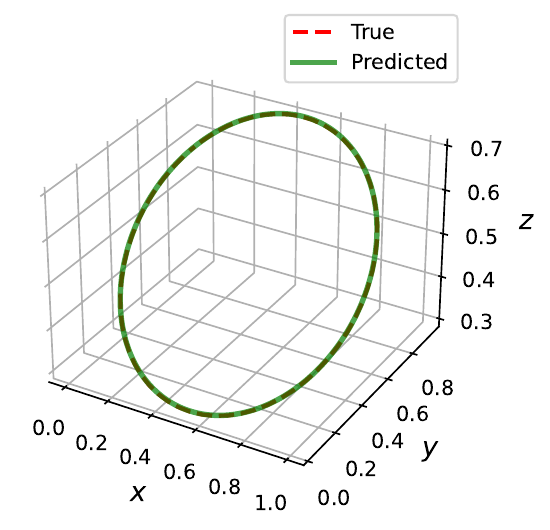}\quad 
	}
	\subcaptionbox{\label{subfig:4b}}
	{
		\includegraphics[width=7cm]{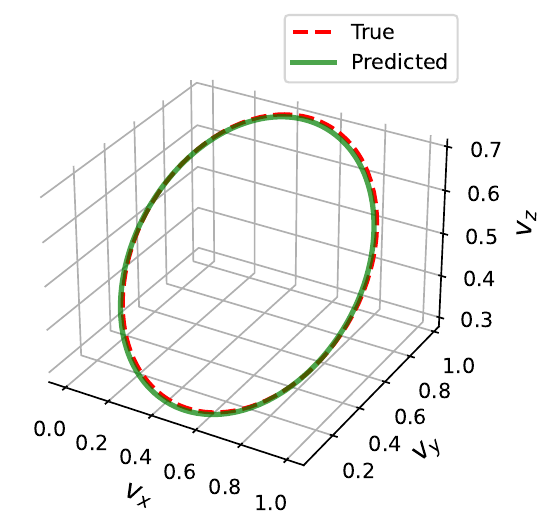}\quad 
		\includegraphics[width=7cm]{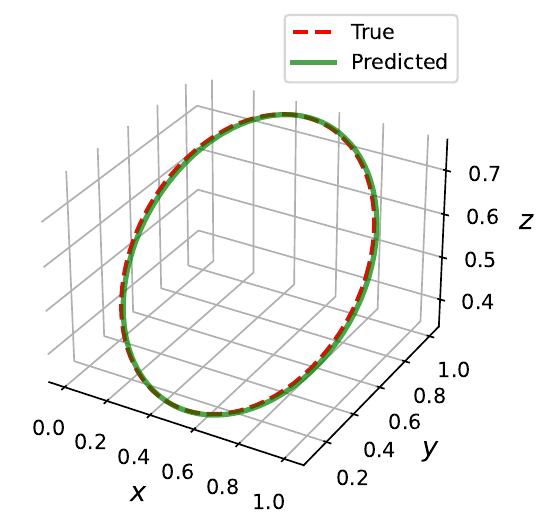}\quad 
	}
	\subcaptionbox{\label{subfig:4c}}
	{
		\includegraphics[width=7cm]{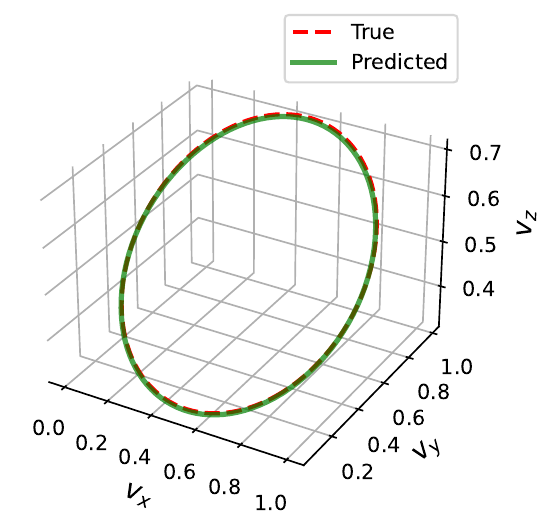}\quad 
		\includegraphics[width=7cm]{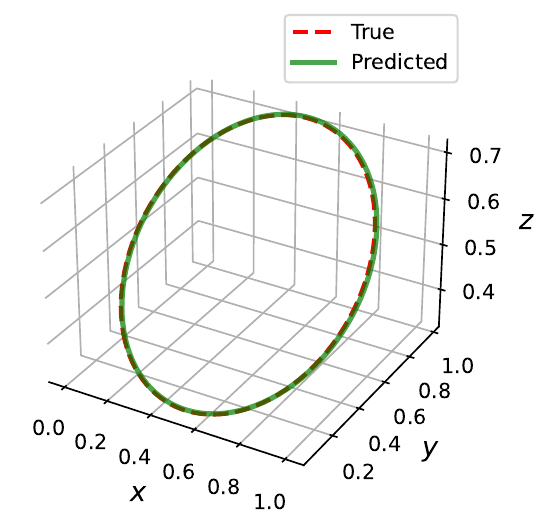}\quad 
	}
	\caption{(Color online) Mutual predictions of velocity and position for planets. The left column is predicted hodograph from position. The right column is predicted orbit from velocity. (a) Predicted resluts of Venus. (b)Predicted resluts of Mars. (c) Predicted resluts of Jupiter.}
	\label{fig4}
\end{figure}
To demonstrate our idea further, we choose two time-dependent series. One is the data of velocity $\vec{v}_{n}$=$(v_{nx},v_{ny},v_{nz})$, where $ v_{nx}$, $v_{ny}$ and $v_{nz}$ are the components of velocity for a planet in heliocentric system. The other is the data of position $\vec{r}_{n}$=$(x_{n},y_{n},z_{n})$, where $x_{n}$, $y_{n}$ and $z_{n}$ are the coordinates of the planet in heliocentric system. We normalize the data of velocity $\vec{v}_{n}$ with constant $V_{max}-V_{min}$, where $V_{max}$  and $V_{min}$ are the maximum and  the minimum of $v_{nx} $. We normalize the data of position $\vec{r}_{n}$ with  constant $R_{max}-R_{min}$, where $R_{max}$  and $R_{min}$ are the maximum  and the minimum of $x_{n} $ for n=1 to 1400, respectively. 

We use the neural network to predict $\vec{v}_{n}$ from $\vec{r}_{n}$. In training set, $\vec{r}_{n}$ (n=1 to 1400)  are the features and $\vec{v}_{n}$ (n=1 to 1400) are the labels. During the prediction, we input $\vec{r}_{n}$ ($n$=1401 to 2600) in test set. And then we output  the  predicted  values of $\vec{v}_{n}$ ($n$=1401 to 2600).

Predicted hodographs of Venus, Mars and Jupiter are shown in the left column of Fig.~\ref{fig4}. The small deviation between the predicted values and  true values implies that the neural network have learned the mapping $\vec{g}$:
\begin{equation}
	\vec{v}_{n} = \vec{g}(\vec{r}_{n}),~~~n=1,2,3...... .\label{g}
\end{equation}
Similarly, we use the neural network to predict $\vec{r}_{n}$ from $\vec{v}_{n}$. The predicted orbits of Venus, Mars and Jupiter are shown in the right column of Fig.~\ref{fig4}. The deviation between the predicted values and  true values is also small. For each of the three planets, we also use  data of 3/4 period in training set to predict data of one period in test set, respectively. And these mutual predictions are similar to Fig.~\ref{fig4}.

The vector mapping (\ref{g}) contains three scalar relations which represent three first integrals for planetary motion in heliocentric system. As we know, planetary motion satisfies the conservation of mechanical energy and conservation of angular momentum. The mechanical energy of the system is expressed as one independent relation
\begin{equation}
	\dfrac{m}{2}v^{2}+ \displaystyle{\frac{k}{r}} =E\label{eq2a},
\end{equation}
where $v$ is the speed of the planet. $r$ is the distance between the planet and the sun, $m$ is mass of the planet. $k$ is the stress of gravitation. $E$ is the mechanical energy.

The angular momentum of the system has three components $L_{x}$, $L_{y}$ and $ L_{z}$. According to poisson theorem, if $L_{x}$ and $L_{y}$  are conserved quantities, $L_{z}$ is also a conserved quantity \cite{HG}. Thus, the conservation of angular momentum merely  contains  two independent relations:
\begin{equation}
	m(yv_{z}-zv_{y})=L_{x}\label{eq2b},
\end{equation}
\begin{equation}
	m(zv_{x}-xv_{z})=L_{y}\label{eq2c},
\end{equation}
where $ v_{x}$, $v_{y}$ and $v_{z}$ are the components of velocity for the planet. $x$, $y$ and $z$ are the coordinates of the planet. Therefore, three independent relations of conserved quantities are satisfied in planetary motion. Eqs.\eqref{eq2a}-\eqref{eq2c} should be equivalent to Eq.\eqref{g}. Since we can solve the relationship between velocity and position from Eqs.\eqref{eq2a}-\eqref{eq2c} in principle. But we do not know the specific correspondence between these relations [Eqs.\eqref{eq2a}-\eqref{eq2c}] and the three first integrals hidden in Eq.\eqref{g} learned by neural networks for planetary motion in heliocentric system.

We infer that the neural network can successfully learn the mapping (\ref{g}) when the number of  first integrals are equal to the dimension of data, instead of learning the differential relationship between velocity and position. To further verify our idea, we discuss an example where the number of conserved quantities is less than the dimension of data in a mechanical system. The planar motion of a particle in a time-dependent central force field can be expressed as
\begin{equation}\label{eq4}
	\dfrac{{\rm d^2 }\vec{r}}{{\rm d}t^2}=-\dfrac{1}{1+t}\dfrac{\vec{r}}{r}\\
\end{equation}
where $\vec{r}$ is position vector. The initial velocity is $\vec{v}_{0}=(0,1)$ and position is $\vec{r}_{0}=(0.5,0)$. There is only one independent conserved quantity (conservation of angular momentum) in this system. 
Using the same neural network and data processing as planetary motion in heliocentric system, we continue to make the mutual prediction of position and velocity. The data set  includes  time-series of velocity and position with the time interval of 0.5. The training set contains data of 6000 samples, and the test set contains data of 4000 samples. The hyperparameters of the neural network are the same as Table~\ref{table2}.

\begin{figure}[ht]
	\centering
	\subcaptionbox{\label{subfig:5a}}
	{
		\includegraphics[width=7cm]{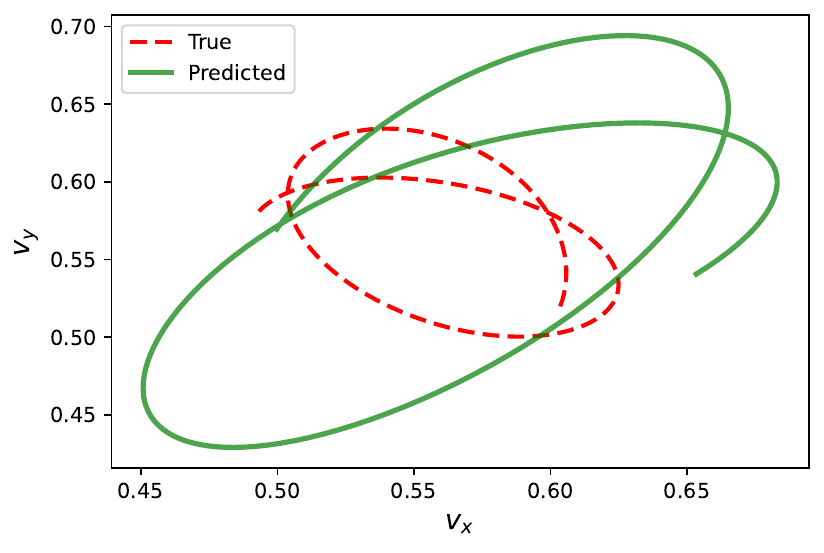}
	}
	\subcaptionbox{\label{subfig:5b}}
	{
		\includegraphics[width=7cm]{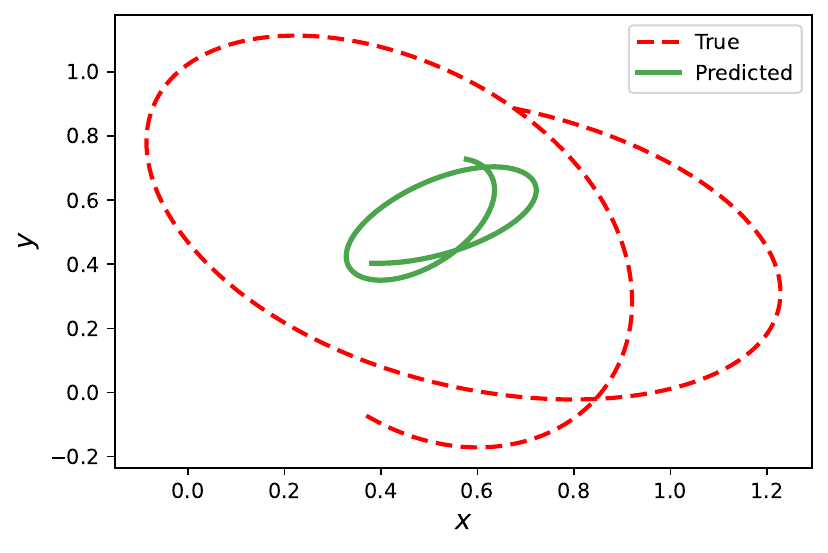}
	}
	\caption{(Color online) Mutual predictions of velocity and position for a particle in a time-dependent central force field. (a) Predicted hodograph from position. (b) Predicted orbit from velocity. }
	\label{fig5}
\end{figure}

The predicted results are shown in Fig.~\ref{fig5}. Obviously, the mutual predictions of position and velocity are very bad. We also obtain the same conclusion for time step 0.1 and 0.01. This example reveals that the neural network we adopt fails in predictions when the number of conserved quantities is less than the dimensions of data, although there is a differential relationship between position and velocity.

These results mentioned above suggest the strong correlation between the existence of conserved quantities in mechanical system and the predicted quality of neural networks. This is the second main result in our paper. Our research provides a new way to explore the existence of conserved quantities in mechanical system  based on neural networks.

\section{\label{sec:level5}Conclusion}

In the above discussion, we make a new attempt to gain physical insights underlying big data by using neural networks. In this sense, the simpler  neural network is, the closer it is likely to the human thought. We have accomplished the goal perfectly with GRU, one of the common neural networks. On the one hand, we find that the precision of predicting orbits in geocentric system strongly depend on whether the data of the Sun is included in training set or not. When considerating the information of the Sun, we make successful prediction of planetary orbits. We fail in prediction without the information of the Sun. Even for Mercury,  we also arrive at the same result. We merely need to change the time interval to 0.001 years, because Mercury has a short period of motion. This striking contrast suggests that the Sun is of great significance in geocentric system and further hints the solar system is heliocentric. On the other hand, we make the mutual prediction of position and velocity in mechanical systems. We demonstrate that the reason for successful prediction is the dimension of system is equal to the number of first integrals. This provides an evidence for the existence of conserved quantities in mechanical system.  We emphasize that our work does not imply the impossibility of learning a dynamic process where there exist no orbitals (e.g., the chaotic systems). For  these systems, the research of learning dynamics from data has been explored, such as the work of Zhao \cite{zhao}.

Liu \cite{liu2021} predicted the orbits of  planets by long short-term memory network, and the results are qualitatively and quantitatively consistent with the present work. We also use fully connected neural networks to explore the existence of conserved quantities and come to the same conclusion. These suggest that our conclusions are of a certain universality. Recently, AI  physicist \cite{Wu} has opened and represents a new research paradigm. Our work contributes a footnote for a special AI physicist. In the future, we hope to clarify the specific correspondence between the three conserved relations [Eqs.\eqref{eq2a}-\eqref{eq2c}] and the three first integrals hidden in Eq.\eqref{g} learned by neural networks for planetary motion in heliocentric system. It will be an exciting mission to use neural networks to discover physical laws directly, which will be a promising road to scientific research.
 
\addcontentsline{toc}{chapter}{Acknowledgment}
\section*{Acknowledgment}
The authors would like to thank Xinran Ma for the helpful discussion in machine learning.

\end{CJK*}  

\addcontentsline{toc}{chapter}{References}
\begin{thebibliography}{99}\footnotesize
\itemsep=-3pt plus.2pt minus.2pt   

\bibitem{nature1943} Jones H S  \href {https://doi.org/10.1038/151573a0}{1943 \emph {Nature} \textbf{151} 573}
\bibitem {nature2015} Lecun Y,  Bengio Y and Hinton G \href{https://doi.org/10.1038/nature14539} {2015 \emph{Nature} \textbf{521} 436}
\bibitem {DL} Goodfellow I, Bengio Y and Courville A 2016 \href{http://www.deeplearningbook.org/} {\emph{Deep Learning}} (Cambridge: MIT Press) 
\bibitem{Caldeira2019}Caldeira J, Wu W L K, Nord B,  Avestruz  C, Trivedi S and Story K T \href{https://www.sciencedirect.com/science/article/pii/S221313371830132X} {2019 \emph{Astronomy and Computing} \textbf{28} 100307}
\bibitem{Yu}Yu H,  Adhikari R X,  Magee  R, Sachdev S  and Chen Y \href{https://link.aps.org/doi/10.1103/PhysRevD.104.062004} {2021 \emph{Phys. Rev. D} \textbf{104} 062004}
\bibitem{Pfeil}Pfeil T, Jordan J, Tetzlaff T, Gr\"ubl A, Schemmel J, Diesmann  M and Meier K \href{https://link.aps.org/doi/10.1103/PhysRevX.6.021023} {2016 \emph{Phys. Rev. X} \textbf{6} 021023}
\bibitem{Pet}Petrovici  M A, Bill J, Bytschok I, Schemmel J and Meier K \href{https://link.aps.org/doi/10.1103/PhysRevE.94.042312} {2016 \emph{Phys. Rev. E} \textbf{94} 042312}
\bibitem{zhaihui}Zhang P, Shen H and Zhai H  \href{https://link.aps.org/doi/10.1103/PhysRevLett.120.066401} { 2018 \emph {Phys. Rev. Lett.} \textbf{120} 066401}
\bibitem{ma2021}Ma X R, Tu  Z C and Ran S J \href{http://cpl.iphy.ac.cn/10.1088/0256-307X/38/11/110301} {2021 \emph{Chin. Phys. Lett.} \textbf{38} 110301}
\bibitem{kau}Kaubruegger R, Pastori L and Budich J C \href{https://link.aps.org/doi/10.1103/PhysRevB.97.195136} {2018 \emph{Phys. Rev. B} \textbf{97} 195136}
\bibitem{sturm}Sturm E J, Carbone M R, Lu D, Weichselbaum A and Konik R M \href{https://link.aps.org/doi/10.1103/PhysRevB.103.245118} {2021 \emph{Phys. Rev. B} \textbf{103} 245118}
\bibitem{yichen}Huang  Y and Moore  J E \href{https://link.aps.org/doi/10.1103/PhysRevLett.127.170601} {2021 \emph{Phys. Rev. Lett.} \textbf{127} 170601}
\bibitem{ZhaoR}Zhao R, Xing B, Mu H, Fu Y and Zhang L \href{http://cpb.iphy.ac.cn/EN/10.1088/1674-1056/ac5d2d} {2022 \emph{Chin. Phys. B} \textbf{31} 056302}
\bibitem{ma2020}Ma X R, He X F and Tu Z C \href{https://doi.org/10.1016/j.engfracmech.2020.107402} {2021 \emph{Engineering Fracture Mechanics}  \textbf{241} 107402}
\bibitem {baldi2016}Baldi P, Cranmer K, Faucett T, Sadowski P and Whiteson D  \href{https://doi.org/10.1140/epjc/s10052-016-4099-4} { 2016 \emph {Eur. Phys. J. C} \textbf{76} 235}
\bibitem{been} Beentjes S V and Khamseh A \href{https://link.aps.org/doi/10.1103/PhysRevE.102.053314}{2020 \emph{Phys. Rev. E}  \textbf{102}  053314}
\bibitem{Giri}Giri S K and Goswami H P \href{https://link.aps.org/doi/10.1103/PhysRevE.99.022104} {2019 \emph{Phys. Rev. E} \textbf{99} 022104}
\bibitem{Rotondo} Rotondo P,  Pastore M and Gherardi  M \href{https://link.aps.org/doi/10.1103/PhysRevLett.125.120601}{2020 \emph{Phys. Rev. Lett.}  \textbf{125}  120601}
\bibitem{huang}Huang H and Goudarzi A \href{https://link.aps.org/doi/10.1103/PhysRevE.98.042311} {2018 \emph{Phys. Rev. E} \textbf{98} 042311}
\bibitem{casert}Casert C,  Vieijra T, Whitelam S and Tamblyn I \href{10.1103/PhysRevLett.127.120602} {2021 \emph{Phys. Rev. Lett.} \textbf{127} 120602}
\bibitem{Liujg}Liu J G, Mao L, Zhang P and Wang L  \href {https://doi.org/10.1088/2632-2153/aba19d}{2021 \emph {Mach. Learn.: Sci. Technol.} \textbf{2} 025011}
\bibitem{zhao}Zhao H  \href{https://doi.org/10.1007/s11433-021-1699-3} {2021 \emph{Sci. China-Phys. Mech. Astron} \textbf{64} 270511 } 
\bibitem{Iten2020}Iten R, Metger T, Wilming H, del Rio L and Renne R   \href{https://link.aps.org/doi/10.1103/PhysRevLett.124.010508} {2020 \emph{Phys. Rev. Lett.}  \textbf{124} 010508}
\bibitem{qin2020}Qin H \href{https://doi.org/10.1038/s41598-020-76301-0}{2020 \emph{Scientific Reports}  \textbf{10}  13929}
\bibitem{gru1}Cho K, van Merri{\"e}nboer B, Gulcehre C, Bahdanau D, Bougares F, Schwenk H and Bengio  Y \href{https://arxiv.org/abs/1406.1078}{2014 arXiv:1406.1078}  [cs.CL]
\bibitem{gru2}Chung J, Gulcehre C, Cho K and Bengio  Y \href{https://arxiv.org/abs/1412.3555} {2014  arXiv:1412.3555}  [cs.NE]
\bibitem{gru3}Cho K, van Merri{\"e}nboer B, Bahdanau D, Bougares F and Bengio  Y \href{https://arxiv.org/abs/1409.1259}{2014 arXiv:1409.1259}  [cs.CL]
\bibitem{de421}\href{https://pypi.org/project/de421/}{https://pypi.org/project/de421/}
\bibitem{rnn1}Elman J L \href{https://doi.org/10.1207/s15516709cog1402_1}{1990 \emph{Cognitive Science}  \textbf{14}  179}
\bibitem{rnn2}Sch{\"a}fer A M and Zimmermann H G  \href{https://doi.org/10.1142/S0129065707001111}{ 2007 \emph{International Journal of Neural Systems} \textbf{1}  253}
\bibitem{keras}Chollet F et al 2015 Keras (\href{https://keras.io.}{https://keras.io.})
\bibitem{tf}Abadi M et al 2015 TensorFlow: Large-scale machine learning on heterogeneous systems software available from tensorflow.org (\href{https://www.tensorflow.org/}{https://www.tensorflow.org/})
\bibitem{adagrad}Duchi J, Hazan E and Singer Y  \href{http://jmlr.org/papers/v12/duchi11a.html}{2011 \emph{Journal of Machine Learning Research} \textbf{12} 2121}
\bibitem{cybenko1989}Cybenko G \href{https://doi.org/10.1007/BF02551274}{ 1989 \emph{Math. Control Signals Systems} \textbf{2}  303}
\bibitem{hornik1989}Hornik K,  Stinchcombe M and White H \href{https://www.sciencedirect.com/science/article/abs/pii/0893608089900208}{ 1989 \emph{Neural Networks } \textbf{2}  359}
\bibitem{HG}Goldstein H  1980 \emph{Classical Mechanics,} 2nd edn. (Massachusetts: Addison Wesley) pp.419
\bibitem{liu2021}Liu Y 2021\emph{ Planetary Orbit Prediction Based on Machine Learning and Exploration of Classical Physics} (M.S. Dissertation) (Beijing: Beijing Normal University) (in Chinese) 
\bibitem{Wu}Wu T and Tegmark M \href{https://link.aps.org/doi/10.1103/PhysRevE.100.033311} {2019 \emph{Phys. Rev. E} \textbf{100} 033311}


\end{thebibliography}
\end{document}